\documentclass[shortAfour,sagev,times]{sagej}

\usepackage{moreverb}
\usepackage{url}
\usepackage[colorlinks,bookmarksopen,bookmarksnumbered,citecolor=red,urlcolor=red]{hyperref}

\usepackage{xcolor}

\newcommand\BibTeX{{\rmfamily B\kern-.05em \textsc{i\kern-.025em b}\kern-.08em
T\kern-.1667em\lower.7ex\hbox{E}\kern-.125emX}}

\def\volumeyear{2016}

\usepackage{xcolor}

\begin{document}

\runninghead{Barth et~al.}

\title{Experimental and numerical investigations on the acoustoelastic effect in hyperelastic waveguides}

\author{Tilmann Barth\affilnum{1}, Natalie Rauter\affilnum{1} and Rolf Lammering\affilnum{1}}

\affiliation{\affilnum{1}Chair of Solid Mechanics, Helmut Schmidt University\,/\,University of the Federal Armed Forces Hamburg, Germany}

\corrauth{Natalie Rauter, Helmut-Schmidt-University / University of the Federal Armed Forces Hamburg, Chair of Solid Mechanics, Holstenhofweg 85, 22043 Hamburg, Germany}
\email{natalie.rauter@hsu-hh.de}

\begin{abstract}
Guided ultrasonic wave based structural health monitoring has been of interest over decades. However, the influence of pre-stress states on the propagation of Lamb waves in thin-walled structures is not fully covered, yet. So far experimental work presented in the literature only focuses on a few individual frequencies, which does not allow a comprehensive verification of the numerous numerical investigations. Furthermore, most work is based on the strain-energy density function by Murnaghan. To validate the common modeling approach and to investigate the suitability of other non-linear strain-energy density functions an extensive experimental and numerical investigation covering a large frequency range is presented here. The numerical simulation comprises the use of the Neo-Hooke as well as the Murnaghan material model. It is found that these two material models show qualitatively similar results. Furthermore, the comparison with the experimental results reveals, that the Neo-Hooke material model reproduces the effect of pre-stress on the difference in the Lamb wave phase velocity very well in most cases. For the $A_0$ wave mode at higher frequencies, however, the sign of this difference is only correctly predicted by the Murnaghan model. In contrast to this the Murnaghan material model fails to predict the sign change for the $S_0$ wave mode.
\end{abstract}


\maketitle

\section{Introduction}

A promising method for the development of reliable structural health monitoring (SHM) systems for lightweight structures utilizes guided ultrasonic waves (GUW). Based on their propagation characteristics and interactions with discontinuities in the waveguide, information about material properties or existing damage in the material can be obtained\cite{Giurgiutiu.2008,Lammering.2018}. In order to analyze the wave propagation data correctly, knowledge about the fundamental wave propagation properties and the inﬂuences of structural properties, damage, inhomogeneities, and environmental conditions is indispensable. Furthermore, the presence of internal stresses and strains leads to a change in the phase velocity of GUW propagating in solids. This phenomenon is known as the acoustoelastic effect\cite{Pao.1985,Qu.1998}. Since all structures are subject to gravity and hence exhibit internal stress states at any time, the interaction of GUW with stress states is crucial for an accurate interpretation of the measurement data.

Fundamental investigations in the field of acoustoelasticity go back to geophysical problems. Biot\cite{Biot1939,Biot1940} developed a theory for the analysis of small deformations and elastic waves under the influence of an internal stress state. Based on this the influence of an internal stress state on homogeneous plane waves can be found in Man and Lu\cite{Man.1987}. In the field of acoustoelasticity it is initially not relevant what causes the internal stress state. However, two different scenarios can be distinguished. If the internal stress state is present in the absence of loads, it is referred to as a residual stress state\cite{Hoger.1986}. Otherwise, it is referred to as a pre-stress state.

More recent studies in the field of acoustoelasticity rely on non-linear constitutive models to describe the occurring effects. In the field of solid mechanics, and especially in the field of GUW based SHM, the application of constitutive models to represent the acoustoelastic effect is limited to the strain-energy density function introduced by Murnaghan\cite{Murnaghan.1951}. This is based on the work of Hughes and Kelly\cite{Hughes.1953}, which deals with the experimental determination of third-order elastic constants by investigating the wave propagation in isotropic materials under the influence of uniaxial and hydrostatic stress states. Thurston and Brugger\cite{Thurston.1964} as well as Toupin and Bernstein\cite{Toupin.1961} extended these investigations to materials with arbitrary crystalline symmetries. Among others Crecraft\cite{CRECRAFT.1967}, Hsu\cite{Hsu.1974} as well as Blinka and Sachse\cite{Sachse.1976} were able to determine the elastic stress states in solids using longitudinal and shear waves. For an overview of the theory of acoustoelasticity and also acoustoplasticity with a focus on hyperelastic anisotropic materials using Murnaghan's constitutive theories, the reader is kindly referred to Pao and Gamer\cite{Pao.1985}, Guz and Makhort\cite{Guz.2000}, and the sources therein.

Early studies on acoustoelasticity in the field of GUW can be found in the works of Hayes\cite{Hayes.1961} and Husson\cite{Husson.1985}, which provide formulations for the modeling of the acoustoelastic effect. Recent literature contains a large number of numerical studies on this topic using a wide variety of geometries and materials\cite{Wilcox.2007,Lematre.2006,Loveday.2008,Mohabuth.2018,Peddeti.2018,Qu.1998,Yang.2019,Yang.2019_2}. However, literature dealing with the validation of numerical and analytical results by experimental data is very limited. Gandhi et al.\cite{Gandhi_Michaels.2012} investigate the acoustoelastic effect by measuring the wave velocities of Lamb waves in biaxially stressed plates and compare the results with analytical results. Qiu et al.\cite{Qiu.2019} investigate the influence on wave propagation for uniaxially pre-stressed plates with a focus on numerical multiphysics simulations considering piezoelectric transducers. Pei and Bond\cite{Pei.2016} deal with the influence of residual stress states on the propagation behavior of Lamb waves and the influence of uniaxial pre-stress states on higher order modes. The listed work is limited to the consideration of the acoustoelastic effect at a few individual frequencies. The authors are not aware of any work on the validation of acoustoelastic behavior over a larger frequency range.

Furthermore, even if the field of acoustoelastic investigations in solids is limited to the strain-energy density function introduced by Murnaghan, a large number of other non-linear constitutive theories are available in the literature. Therefore, in this work the Neo-Hooke material model is additionally investigated, which goes back to the work of Ronald Rivlin\cite{belytschko.2013,Ogden.2013} and represents a classical extension of Hooke's law to finite deformation elasticity. Further constitutive models for the representation of initial stress states were primarily developed within the field of biomechanics, where the focus is on residual stresses. Since, the present work focuses on the inﬂuence of a pre-stress state, the reader is kindly referred to the work of Ogden\cite{Ogden.2011,Ogden.2013,ogden.2010}  and Hoger\cite{Hoger.1986,Hoger.1993} and the sources contained therein for more details about residual stresses.

Based on this the present study holds a comprehensive investigation of the influence of an uniaxial pre-stress state on the GUW propagation in an isotropic waveguide. Therefore, first experiments are conducted to observe the acoustoelastic effect over a large frequency-thickness range of up to 3~MHzmm. For the data acquisition and evaluation a technique is used which was introduced by the authors in previous work\cite{Barth.SMS.2023,Barth.ExpM.2022}. In a second step numerical simulations are carried out. Here not only the material model of Murnghan but also the well-known Neo-Hooke material model is used to incorporate the acoustoelastic effect. Using a small segment of the pre-stressed waveguide and following an approach presented by the authors in previous work\cite{Barth.GAMM.22.FML}, again the acoustoelastic effect is studied over a frequency-thickness range of up to 3~MHzmm for both material models. To evaluate the numerical simulation and hence, the suitability of the different material models, the obtained numerical data is compared with the experimental results in a final step.

Subsequently, the structure is as follows. First, the theoretical fundamentals of the acoustoelastic effect are presented. Next are the experimental investigations on how pre-stress states influence the wave propagation properties in an aluminum strip specimen. This is followed by numerical simulations. Afterwards, the experimentally obtained data is compared to the results of the numerical simulations based on the Neo-Hooke and Murnaghan material models. The presented work closes with a brief summary and conclusion.

\section{Fundamentals}
\label{sec:fundamentels}
Investigations on pre-stressed hyperelastic structures require a revisiting of the theoretical foundations, such as the balance of linear momentum and the constitutive equations. This is the subject of this section.

\subsection{Kinematics}
\begin{figure}[hbt]
	\centering
    \def\svgwidth{\columnwidth}
    \includegraphics{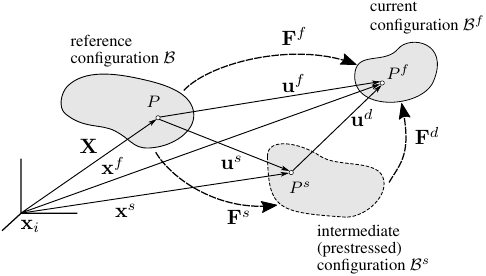}
	\caption{Sketch of configurations, position vectors, displacements and deformation gradients.}
	\label{fig:configurations}
\end{figure}

The starting point of the investigations is the decomposition of the total deformation of the body $\mathcal B$ into two parts, namely the deformation due to the pre-stress and the deformation due to the propagation of the elastic wave \cite{Peddeti.2018}. The pre-stressing transfers the body $\mathcal B$ from the reference configuration into the intermediate configuration $\mathcal B^s$. The subsequently propagating elastic waves in the pre-stressed body lead to the current configuration $\mathcal B^f$. A sketch of this decomposition is shown in Figure \ref{fig:configurations}. Mathematically, this decomposition is described by a multiplicative split of the deformation gradient
\begin{equation}
\mathbf F^f = \mathbf F^d \cdot \mathbf F^s \ .
\label{eq:f1}
\end{equation}
Here, $\mathbf F^f$, $\mathbf F^d$, and $\mathbf F^s$ denote the deformation gradients of the total deformation, the deformation due to the elastic waves and the pre-stress, respectively.
\subsection{Balance of linear momentum}
Next, the balance of linear momentum in the current configuration $\mathcal B^f$  is expressed as
\begin{equation}
\operatorname{Div} \mathbf P^f = \rho_0 \frac{\partial^2 \mathbf u^f}{\partial t^2} \, . 
\label{eq:f2}
\end{equation}
Here, $\mathbf P$ denotes the first Piola-Kirchhoff stress tensor, $\rho_0$ the density in the reference configuration, $\mathbf u$ the displacement vector and $t$ the time. The superscript $f$ stands for the current configuration and $\operatorname {Div}$ denotes the divergence operator with respect to the reference configuration $\mathcal B$. Introducing the second Piola-Kirchhoff stress tensor $\mathbf S^f$ one obtains
\begin{align}
\mathbf P^f 	&= \mathbf F^f \cdot \mathbf S^f  \notag \\
			&= \mathbf S^f + \operatorname{Grad} \mathbf u^f \cdot \mathbf S^f \, ,
\label{eq:f3}
\end{align}
where $\mathbf F^f = \mathbf I + \operatorname{Grad} \mathbf u^f$ holds, $\mathbf I$ is the unit tensor and $\operatorname{Grad} \mathbf u^f$ the gradient of the displacement vector $\mathbf u^f$. Equation\,\eqref{eq:f2} is now rewritten as 
\begin{equation}
\operatorname{Div} \lbrace \mathbf S^f + \operatorname{Grad} \mathbf u^f \cdot \mathbf S^f \rbrace = \rho_0 \frac{\partial^2 \mathbf u^f}{\partial t^2} \, . 
\label{eq:f4}
\end{equation}
Analogously, the equation of equilibrium in the intermediate configuration $\mathcal B^s$ reads
\begin{equation}
\operatorname{Div} \lbrace \mathbf S^s + \operatorname{Grad} \mathbf u^s \cdot \mathbf S^s \rbrace = \rho_0 \frac{\partial^2 \mathbf u^s}{\partial t^2} \, . 
\label{eq:f5}
\end{equation}
Subtraction\cite{Peddeti.2018} of Equation\,\eqref{eq:f5} from Equation\,\eqref{eq:f4} yields
\begin{align}
& \operatorname{Div} \lbrace \mathbf S^f - \mathbf S^s + \operatorname{Grad} \mathbf u^f \cdot \mathbf S^f - \operatorname{Grad} \mathbf u^s \cdot \mathbf S^s \rbrace \notag \\
& = \rho_0 \frac{\partial^2 (\mathbf u^f - \mathbf u^s)}{\partial t^2} \, . 
\label{eq:f6}
\end{align}
Now $\mathbf S^d$ is introduced as the difference between the stresses in the current and intermediate configuration
\begin{equation}
\mathbf S^d = \mathbf S^f - \mathbf S^s \, 
\label{eq:f7}
\end{equation}
and $\mathbf u^d$ as the respective displacement difference
\begin{equation}
\mathbf u^d = \mathbf u^f - \mathbf u^s \, .
\label{eq:f8}
\end{equation}
After some rearrangements and by use of $\mathbf P^d = \mathbf F^f \cdot \mathbf S^d$ one finally obtains
\begin{equation}
\operatorname{Div} \lbrace \mathbf P^d + \operatorname{Grad} \mathbf u^d \cdot \mathbf S^s \rbrace 
= \rho_0 \frac{\partial^2 \mathbf u^d}{\partial t^2} \, ,
\label{eq:f9}
\end{equation}
which is considered as the strong form of equilibrium of a body under pre-stress $\mathbf S^s$. It should be noted, that the well known formulation for the stress-free body is obtained for $\mathbf S^s = \mathbf 0$.
\subsection{Hyperelastic material models}
In this work, it is assumed that the structures under consideration can be described by a hyperelastic material law. Then, a strain-energy density function $W$, which depends on the deformation gradient $\mathbf F^f$ and the pre-stress $\mathbf S^s$, is defined per unit volume in $\mathcal B$. From $W$, the stress tensor and the elasticity tensor is derived by the first and the second derivative with respect to the deformation gradient $\mathbf F$, respectively. 

When analyzing acoustoelastic phenomena, the Murnaghan material model \cite{Murnaghan.1937,Murnaghan.1951} is often used, see e.g. \cite{Husson.1985,Abbasi.2016, Peddeti.2018}. In terms of the invariants $I_1, I_2$, and $I_3$ of the right Cauchy-Green tensor $\mathbf C$ the Murnaghan strain-energy density function $W^{\text M}$ reads
\begin{align}
W^{\text M} = & \frac{\lambda}{8} (I_1 - 3 )^2 + \frac{\mu}{4} ( I_1^2 - 2 I_1 - 2 I_2 + 3) \notag \\
& + \frac{\ell}{24} (I_1 - 3 )^3 + \frac{m}{12}(I_1 - 3)(I_1^2 - 3 I_2) \notag \\
& + \frac{n}{8}(I_1 - I_2 + I_3 - 1) \, .
\label{eq:M}
\end{align}
Here, $\lambda$ and $\mu$ are the Lam\'e constants sometimes also referred to as the second-order elastic constants and $\ell,\,m$, and $n$ the third-order constants of the material model. Equation\,\eqref{eq:M} leads to an extension of the classical theory of elasticity and is referred to as second order theory \cite{Rivlin.1953}. It becomes clear that this material model requires five constants, of which those of third order are difficult to determine.

A more easily applicable strain-energy function, often used in nonlinear elasticity, requiring only the both Lam\'e constants, is associated with the Neo-Hooke material model. In case of a compressible material $W^{\text{NH}}$ is written as \cite{belytschko.2013} 
\begin{equation}
W^{\text{NH}} = \frac12 \lambda \, (\operatorname{ln} I_3)^2 - \mu \,\operatorname{ln} I_3 + \frac 12 \mu \, ( I_1 - 3 ) \, .
\label{eq:NH}
\end{equation}
In this work, the use of the Neo-Hooke material model is motivated by the fact, that the pre-stress may cause a finite deformation moving the reference configuration to the intermediate configuration and resulting in a deformation dependent elasticity tensor. The subsequent infinitesimal deformation generated by the elastic waves, however, can be analyzed by a linearized theory in the intermediate configuration.

In the literature, see e.g. \cite{Ogden.2013}, further hyperelastic material models can be found, see e.g. those according to Mooney-Rivlin or Yeoh. However, investigations on these models are beyond the scope of this work.
\section{Experimental investigations}
\label{sec:Experiments}

\begin{figure*}[ht]
  \centering
  \includegraphics[scale=1]{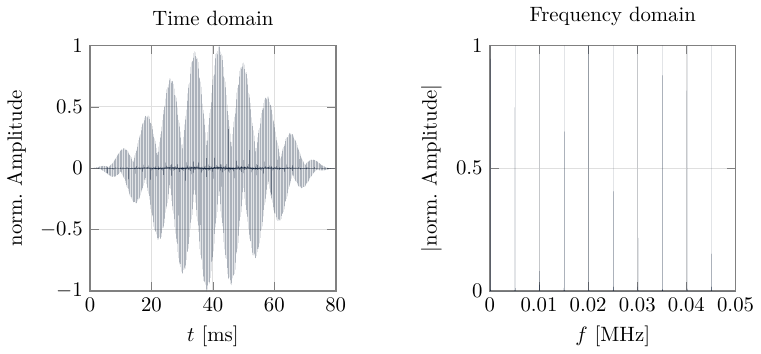}
  \caption{Multifrequent excitation signal.}
  \label{fig:Anregung}
\end{figure*}

This section deals with the experimental investigation of the acoustoelastic effect in a pre-stressed aluminum strip specimen. Therefore, first the experimental setup is described before presenting the obtained results. With this data at hand, the numerical modeling approach, presented in the subsequent section, is validated in a last step by comparing it with the measurement data.

\subsection{Methodology and procedure}

For the observation of the GUW propagation the scanning laser vibrometry is a suitable 
technique.\cite{Koehler.2006,Schpfer.2013,Zimmermann.2018} This allows one to measure the velocity component of the wave field that is parallel to the laser beam on the surface of a specimen. The result is a velocity matrix $\mathbf{v}(x_1,x_2,t)$ as a function of the spatial coordinates $x_1, x_2$ and time $t$. Since GUW are multimodal and dispersive by nature\cite{Lamb.1917,graff.2012} a non-uniform 2D-DFT related method is used for the data acquisition and postprocessing. The 2D-DFT was first introduced by Alleyne et al.\cite{Alleyne.1991}. It shows the suitability of the 2D-DFT for the dispersion relation extraction. The authors adapted the original approach in previous work\cite{Barth.ExpM.2022,Barth.SMS.2023} to ensures high level accuracy and automated determination of the GUW dispersion behavior over a large frequency range. For more information about the 2D-DFT the reader is kindly referred to common literature\cite{Hora.2012,Su.2009}.

To be able to analyze the effect of a pre-stress state on the wave propagation by also addressing the multimodal and dispersive nature of GUW the postprocessing procedure focuses on the frequency-wavenumber domain. This also allows the use of a multifrequent excitation signal. Therefore, a superposition of sinusoidal oscillations is applied in this work. To avoid any influence by a finite sinus burst a Hann window is added \cite{blackman.1959}. The distance between the excited frequencies is set to 5~kHz. By repeating the measurement 40 times with an excitation signal shifted by 0.125~kHz the frequency resolution is significantly increased. Figure~\ref{fig:Anregung} gives an exemplary excitation signal in the time and frequency domain. It covers a frequency range from 0.125~kHz to 995.25~kHz with steps of 5~kHz and has a total length of 80~ms. To illustrate the frequency resolution more in detail the frequency range on the right side in Figure~\ref{fig:Anregung} is reduced. Details regarding the extraction of the frequency - wavenumber pairs and the corresponding peak-search algorithm as well as the identification of outliers can be found in previous work by the authors \cite{Barth.ExpM.2022,Barth.SMS.2023}.

Within this work the influence of a pre-stress state on the GUW propagation in aluminum is analyzed in a range from 0~MPa up to 100~MPa. Following this the wave propagation is measured first in an unloaded specimen before increasing the external load by 10~MPa steps until a stress state of 100~MPa is reached. For each configuration the measurement is repeated 40 times following the definition of the excitation signal. Furthermore, with respect to the designated wavenumber range a maximal excitation frequency of 1~MHz and 2~MHz is used, respectively.

\subsection{Setup and specimen definition}
\label{sec:Specimens}

\begin{figure}[ht]
  \centering
  \def\svgwidth{\columnwidth}
  \includegraphics{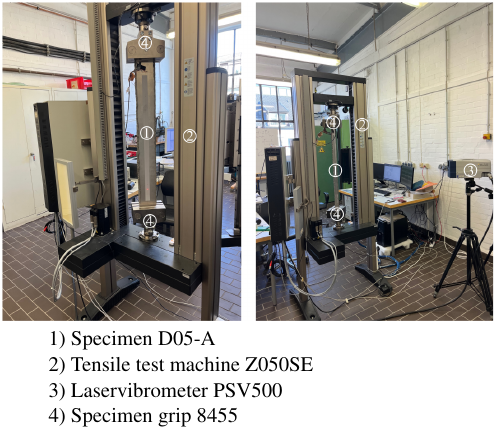} 
  \caption{Experimental setup for the laser vibrometer measurement of the wave propagation in a pre-stressed specimen.}
  \label{fig:Messaufbau}
\end{figure}

To measure the wave propagation in pre-stressed specimens the waveguide is mounted vertically in a tensile test machine Z050SE as depicted in Figure~\ref{fig:Messaufbau}. The excitation signal is first amplified by an high voltage amplifier before it is passed to a piezo-electric element, which is bonded to the surface of the specimen and excites the wave field. To capture the wave propagation the laser vibrometer is aligned so that the laser beam is perpendicular to the specimen at the midpoint of the measurement path. Due to this the measurement signal represents dominantly the out-of-plane velocity component of the wave propagation. Therefore, the frequency range for the $S_0$ wave mode is selected in such a way to ensure prominent out-of-plane displacement fields.

\begin{figure}[htb]
  \centering
  \includegraphics[width=\columnwidth]{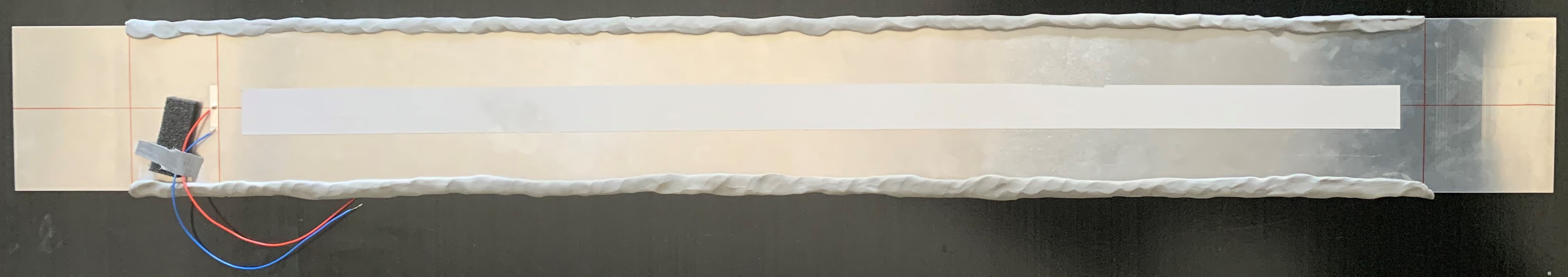}
  \caption{Strip specimen made from AlMg3 with a size of 1100~mm $\times$ 110~mm $\times$ 0.5~mm.}
  \label{fig:Specimen}
\end{figure}

\begin{table*}
\small\sf\centering
\caption{Specimen specification.}
\label{tab:Specimen}
\begin{tabular}{llcccccccc}
\toprule
& Parameter & Unit & D2-A & D2-B & D2-C & D2-D & D05-A & D05-B & D05-C\\
\midrule
\multicolumn{10}{l}{Geometry} \\
\midrule
$d$ & Thickness  & mm & 2 & 2 & 2 & 2 & 0.5 & 0.5 & 0.5 \\
$\ell$ & Length  & mm & 1100 & 1100 & 1100 & 1100 & 1100 & 1100 & 1100 \\
$w$ & Width  & mm & 110 & 110 & 110 & 110 & 110 & 110 & 110\\
\midrule
\multicolumn{10}{l}{Measurement setup} \\
\midrule
& Distance specimen - vibrometer & mm & 557 & 556 & 1160 & 1158 & 1159 & 1161 & 1161 \\
$\ell_{\text{mes}}$ & Length measurement path  & mm & 432 & 432 & 669 & 669 & 669 & 669 & 667 \\
$\Delta \ell$ & Distance measurement points  & mm & 0.86 & 0.86 & 1.33 & 1.33 & 1.33 & 1.33 & 1.33 \\
$f_{\text{mes}}$ & Sampling rate & MHz & 6.25 & 6.25 & 6.25 & 6.25 & 6.25 & 6.25 & 6.25 \\
$t_{\text{mes}}$ & Measurement time & ms & 80 & 80 & 80 & 80 & 80 & 80 & 80\\
$f_{\text{max}}$ & Maximal excitation frequency & MHz & 2 & 2 & 1 & 1 & 1 & 1 & 1\\
\midrule
\multicolumn{10}{l}{Evaluation boundaries} \\
\midrule
min. $\Tilde{\nu}d$ & Minimal wavenumber & mm/m & 92.56 & 92.56 & 59.79 & 59.79 & 14.95 & 14.95 & 14.76 \\
max. $\Tilde{\nu}d$ & Maximal wavenumber & mm/m & 1163 & 1163 & 751.88 & 751.88 & 187.67 & 187.67 & 187.97 \\
min. $fd$ & Minimal frequency & kHzmm & 1500 & 1500 & 1.47 & 1.47 & 1.47 & 30 & 30 \\
max. $fd$ & Maximal frequency & kHzmm & 3454 & 3454 & 284 & 284 & 284 & 1936 & 1936 \\
\bottomrule
\end{tabular}
\end{table*}

Furthermore, the size of the grips limit the specimen width. To ensure a mostly homogeneous stress field within the specimen the grip size of 110~mm is identical to the width of the specimen. Due to the use of a non-uniform 2D-DFT for the data extraction and an additional damping on the specimens edges by plasticine the effect of edge reflections can be reduced significantly\cite{Barth.SMS.2023}. The length of the specimen is limited by the test area height. This leads to a specimen size of 1100~mm $\times$ 110~mm. Figure~\ref{fig:Specimen} depicts a strip specimen made from aluminum AlMg3. Based on the clamping length of the grips and the area required for the piezo-electric element the maximum measurement length is 700~mm.

Lastly the specimen thickness is directly linked to the desired frequency and wavenumber range of the obtained dispersion diagrams by the dispersion relation. Due to the Nyquist-criterion with respect to time and space as well as the maximum resolution of the measurement path the observable wavenumber and frequency range are limited \cite{Shannon.1949}.

The minimal distance between two measurement points determines the maximum wavenumber. Following the Nyquist-criterion the highest wavenumber is given by half the spatial sampling rate \cite{Barth.SMS.2023}. Furthermore, in previous work of the authors it was shown that there shall be twenty full wave cycles along the measurement distance \cite{Barth.ExpM.2022}, which gives the minimum wavenumber. Following this the minimal and maximal wavenumbers can be obtained by
\begin{equation}
    \text{min.} \ \Tilde{\nu}d = \frac{20}{\ell_{\text{mes}}}d
\end{equation}
and
\begin{equation}
    \text{max.} \ \Tilde{\nu}d = \frac{1}{2\Delta \ell}d,
\end{equation}
respectively. The variables are defined in Table~\ref{tab:Specimen}.

Based on the obtained wavenumber boundaries the corresponding frequency range can be derived by solving the analytical framework of GUW. With a specimen of 2~mm thickness the $S_0$ wave mode can be analyzed sufficiently accurate for frequencies between 1.5~MHzmm and approximately 3.45~MHzmm. With respect to the same boundary conditions and the identical thickness of the specimen the dispersion diagram for the $A_0$ wave mode is evaluable in the frequency range from 1.47~kHzmm to 284~kHzmm. In comparison to the $S_0$ wave mode the $A_0$ wave mode is only sufficiently measurable at significant lower frequencies. Due to this the propagation properties of the $A_0$ wave mode is investigated utilizing specimens of different thicknesses. Beside specimens of 2~mm thickness also specimens of 0.5~mm thickness are used. Reducing the thickness by a factor of 4 enables one to increase the maximal frequency to 1.93~MHzmm. In total, there are two specimens with a thickness of 2~mm and three specimens with a thickness of 0.5~mm. An overview of the specimens and the corresponding wavenumber thickness and frequency thickness ranges, respectively, are provided in Table~\ref{tab:Specimen}. The wave propagation within the specimens with a thickness of 2~mm are measured twice with different distances between the specimen's surface and the vibrometer. To ensure clarity about the experimental setup these configurations are added separately to Table~\ref{tab:Specimen}. Therefore, specimens D2-A and D2-C as well as D2-B and D2-D are identical specimens.

\subsection{Results and discussion}
\label{sec:Exp_Res}

\begin{figure*}[htb]
    \centering
    \def\svgwidth{\columnwidth}
    \includegraphics{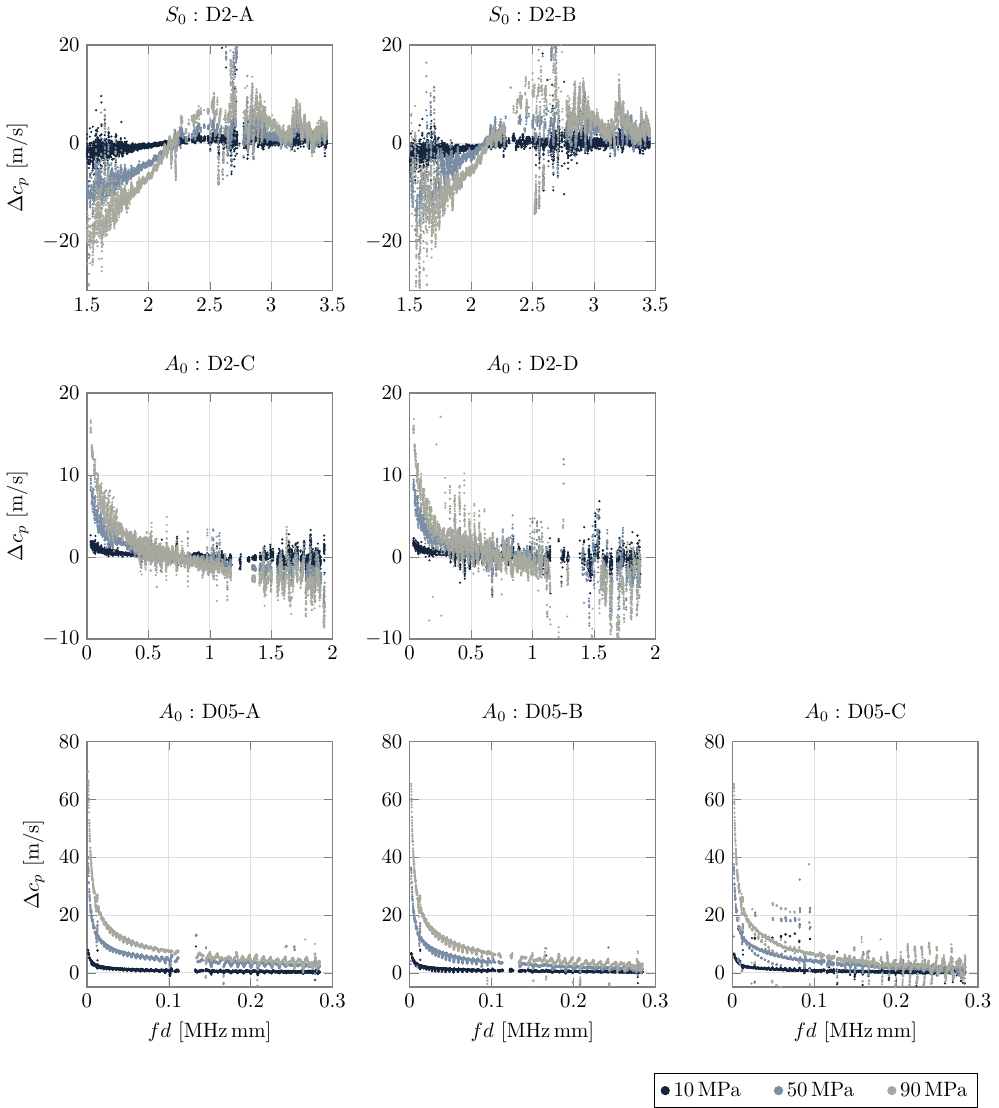} 
  \caption{Phase velocity difference $\Delta c_p$ for the $S_0$ and $A_0$ wave modes at different stress levels.}
  \label{fig:cp_diff_exp_S0_A0}
\end{figure*}

In this section the results of the experimental investigations are presented. To analyze the influence of pre-stress states on the wave propagation the difference of the phase velocity at certain frequencies is derived from the measurement data for each load step. The phase velocity difference $\Delta c_p$ is defined as
\begin{equation}
	\Delta c_p =  c_p^{\text{HS}} - c_p^{\text{LS}},
	\label{eqn:delta_cp}
\end{equation}
where the index $\text{HS}$ indicates the higher stress level and \text{LS} the lower stress state, respectively. For the results presented below for the experimental investigation $c_p^{\text{LS}}$ refers to the 10~MPa load step. An identical definition is used for the stress state difference
\begin{equation}
	\Delta \sigma =  \sigma^{\text{HS}} - \sigma^{\text{LS}} \, .
	\label{eqn:sigma}
\end{equation}

For the sake of clarity, Figure~\ref{fig:cp_diff_exp_S0_A0} only holds the change of the phase velocity for the $S_0$ and $A_0$ wave modes for stress differences of 10~MPa, 50~MPa, and 90~MPa. As mentioned before, the results for the $A_0$ wave mode at low frequencies are based on the specimens A05-A, A05-B, and A05-C. Whereas at higher frequencies the results belong to the specimens A2-C and A2-D, see also Table~\ref{tab:Specimen}. 

The diagrams for both wave modes and all specimens indicate similar behavior. First, the measurements of the wave propagation in different specimens lead to comparable results. For the $S_0$ wave mode the measurements based on D2-A and D2-B reveal both a decrease of the phase velocity within the frequency range of up to 2.15~MHzmm. At about 2.15~MHzmm the sign of the phase velocity difference $\Delta c_p$ changes from negative to positive. It is worth noticing that the sign change is identical for both specimens and for all load cases. Therefore, this point appears to be a significant property of the acoustoelastic effect with respect to the given material layout and load specification. Afterwards the phase velocity difference increases further until it reaches its maximum at about 2.5~MHzmm. Finally it drops with increasing frequency. However, there is no further sign change.

Considering the influence of a stress state on the $A_0$ wave mode, first, the results based on the specimens D05-A, D05-B, and D05-C are discussed. Here, the measurement data reveals again the influence of different stress states on the phase velocity. In contrast to the $S_0$ wave mode the phase velocity difference decreases with an increasing frequency. However, there is no sign change. Furthermore, in contrast to the $S_0$ wave mode the influence of a pre-stress on the wave propagation is even more prominent especially in the low frequency domain. Keeping in mind that the phase velocity of the $A_0$ wave mode is far below the $S_0$ wave mode at low frequencies, this effect appears to be significant for the $A_0$ wave mode at low frequencies.

The results are supported by the observation of the $A_0$ wave mode based on the specimens D2-C and D2-D at higher frequencies. Here again the decreasing development of the phase velocity diference is observable. Due to the larger frequency domain the data also reveals a sign change between 0.7~MHzmm and 0.8~MHzmm. Due to the very small influence on the phase velocity the exact value cannot be determined.  

\begin{figure*}
    \centering
    \def\svgwidth{\columnwidth}
    \includegraphics{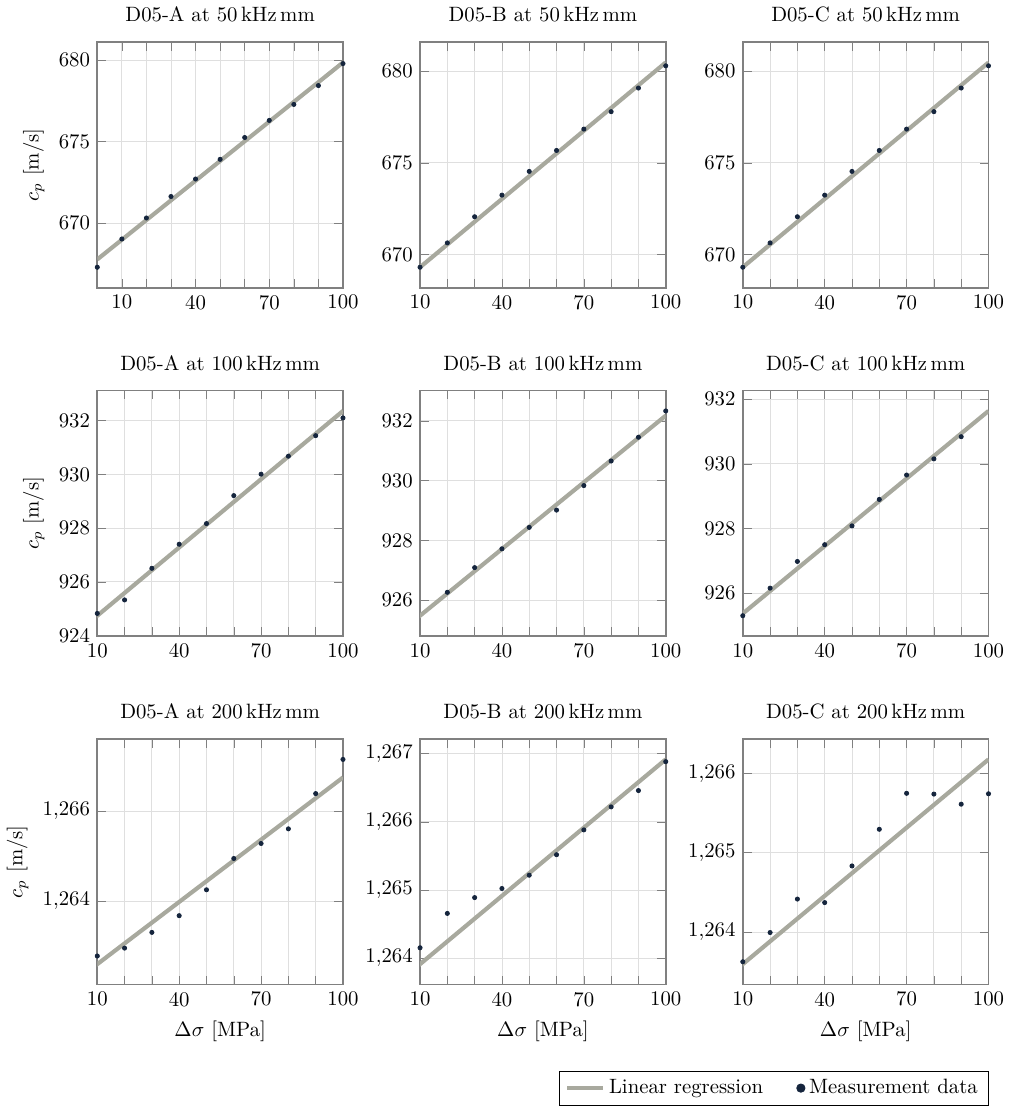} 
  \caption{Approximation of the phase velocity $c_p$ at 50~kHz, 100~kHz, and 200~kHz by a linear regression for the $A_0$ wave mode, obtained from the specimens D05-A, D05-B, and D05-C.}
  \label{fig:Linear_Verlauf_exp}
\end{figure*}

Based on the presented data it is the authors' intention to analyze the correlation between the phase velocity difference and the pre-stress state more in detail. Thereby, special emphasis is laid on the $A_0$ wave mode at low frequencies, because the acoustoelastic effect appears to be most prominent here. For the following discussion about the course of the phase velocity difference due to a pre-stess state, the individual experimental phase velocities are plotted over the applied stress difference in Figure~\ref{fig:Linear_Verlauf_exp}. Furthermore, a linear regression based on these measurement results is added to the plots. 

The first row of Figure~\ref{fig:Linear_Verlauf_exp} shows the comparison between the measured data and the regression curve for a frequency-thickness product of 50~kHz. The plots  clearly indicate a linear relation between an increasing load state and an increase of the phase velocity for all three specimens. In the second row, this relationship can also be clearly concluded for a frequency-thickness product of 100~kHz. However, with an increasing frequency the deviation between the measurement data and the linear regression increases. This can be observed in the last row, where the data is depicted for a frequency-thickness product of 200~kHz. Although the data still indicate a linear regression, this can no longer be clearly confirmed. This conclusion is mostly driven by the decreasing phase velocity difference due to a pre-stress state at higher frequencies as shown in Figure~\ref{fig:cp_diff_exp_S0_A0} and is also linked to an decreasing value of $R^2$ for the corresponding curve fits.   

\section{Numerical investigations}
The finite element method is used for numerical analysis of the dynamically loaded pre-stressed specimens. However, before the computations are carried out, the weak form of equilibrium has to be formulated from Equation \eqref{eq:f9} and then linearized.

The computation of wave propagation in pre-stressed structures is carried out in a two-step procedure. First, the structure is loaded by external forces or a prescribed displacement, resulting in internal pre-stresses and accompanied with a mapping from the reference to the intermediate configuration, see Figure \ref{fig:configurations}. Subsequently, the dynamical loads which are applied harmonically for the generation of the elastic waves are considered. They cause a mapping from the intermediate to the current configuration. To determine dispersion diagrams a linear eigenvalue problem is solved.
\subsection{Weak form}
The weak form of equilibrium is obtained by multiplying Equation\,\eqref{eq:f9} with the virtual displacement vector $\delta \mathbf u^d$ and subsequent integration
\begin{align}
& \int\limits_{\mathcal B} \operatorname{Div} \mathbf P^d \cdot \delta \mathbf u^d \text{d}V 
+ \int\limits_{\mathcal B} \operatorname{Div} \lbrace \operatorname{Grad} \mathbf u^d \cdot \mathbf S^s \rbrace \cdot \delta \mathbf u^d \text{d}V \notag \\
& = \int\limits_{\mathcal B} \rho_0 \frac{\partial^2 \mathbf u^d}{\partial t^2} \delta \mathbf u^d \text{d}V \, .
\label{eq:f10}
\end{align}
Further straightforward rearrangements by use of the product rule and the divergence theorem yield
\begin{align}
&\int\limits_{\mathcal B} \mathbf F^f \! \cdot \mathbf S^d \colon \negthickspace \operatorname{Grad} \delta \mathbf u^d \text{d}V 
+ \negmedspace \int\limits_{\mathcal B} \operatorname{Grad} \mathbf u^d \! \cdot \mathbf S^s \colon \negthickspace \operatorname {Grad} \delta \mathbf u^d \text{d}V \notag \\
&=\int\limits_{A_\sigma} \lbrace \mathbf F^f \! \cdot \mathbf S^d \! \cdot \mathbf N + \operatorname{Grad} \mathbf u^d \cdot \mathbf S^s \! \cdot \mathbf N \rbrace \! \cdot \! \delta \mathbf u^d \text{d}A \notag \\
&+ \int\limits_{\mathcal B} \rho_0 \frac{\partial^2 \mathbf u^d}{\partial t^2} \delta \mathbf u^d \text{d}V \, .
\label{eq:f11}
\end{align}
The two expressions on the left-hand side represent the virtual work of the internal stresses, while the virtual work of the external and the inertia loads appear on the right-hand side. Again, the standard formulation for the stress-free body is obtained for $\mathbf S^s = \mathbf 0$. 

\subsection{Linearization}
Equation \eqref{eq:f11} is nonlinear with respect to the displacement vector $\mathbf u$ after the implementation of the nonlinear hyperelastic constitutive equation, so that it has to be linearized with regard to the finite element analyses. The linearization procedure is presented e.\,g. in \cite{Vu.2020} or directly carried out in some finite element codes, see e.\,g. \cite{comsol.2015}, for both constitutive models which are presented above and is not shown here for sake of brevity.

Discretization and introduction of the standard finite element notation results in the following matrix equation for the entire system
\begin{align}
    [\mathbf M][\ddot{\mathbf u}_j] + 
    \big[ [\mathbf K_d] + [\mathbf K_s] \big] [{\mathbf u}_j ]
    = [\mathbf F] \,.
\label{eq:f12}
\end{align}
Here, $[\mathbf M]$ stands for the global mass matrix, $[\mathbf K_d]$ and $[\mathbf K_s]$ for the linearized global stiffness matrix and the initial stress matrix, respectively, and the sum of both represents the tangent stiffness matrix. The column matrix $[\mathbf u_j]$ includes the unknown displacements of the system. On the right-hand side, $[\mathbf F]$ denotes the the column matrix of external loads. It should be noted, that damping is not included in this model.

\subsection{Eigenvalue problem}

The following considerations are made with regard to isotropic strip-shaped waveguides, see Figure~\ref{fig:Specimen}, but can easily adapted to other specimens. A corresponding model is shown in Figure~\ref{fig:num_modell}. The nodal displacements $\mathbf{u}_j$ due to the propagating waves can be assumed as
\begin{equation}
	\mathbf{u}_j = \hat{\mathbf{u}}_j e^{ikx_1}e^{-i \omega t} \, .
	\label{eq:f13}
\end{equation}
Here, $\hat{\mathbf{u}}_j$ is the wave amplitude, $k$ the circular wavenumber, $\omega$ the circular frequency, $x_1$ the propagation direction, $t$ the time, and $i=\sqrt{-1}$. It becomes obvious, that the displacements are harmonic with respect to the $x_1$ axis. Now, a unit cell of length $\Delta x_1$ of the waveguide is considered that can be imagined repeatedly in the direction of wave propagation, see Figure~\ref{fig:num_modell}a. Equation \eqref{eq:f13} is now used to establish the following relationship between the nodal displacements at the left (index $l$) and right (index $r$) boundary of the unit cell
\begin{equation}
	\mathbf{u}_j^r = e^{ik\Delta x_1} \mathbf{u}_j^l \, .
	\label{eq:f14}
\end{equation} 
In this way, Floquet boundary conditions are formulated \cite{Hakoda.2018,Gomez.2015}, which become conventional periodic boundary conditions for $\Delta x_1 = 2\pi/k$ resulting in $\mathbf u_k^l = \mathbf u_k^r$ \cite{sorohan.2011}. Furthermore, the nodal forces at both ends are related by
\begin{equation}
	\mathbf{F}_j^r = -e^{ik\Delta x_1} \mathbf{F}_j^l \, .
	\label{eq:f15}
\end{equation} 
Implementation of Equations \eqref{eq:f14} and \eqref{eq:f15} into Equation \eqref{eq:f12} and some subsequent rearrangements define the matrices $[\bar{\mathbf K}_d]$, $[\bar{\mathbf K}_s]$, and $[\bar{\mathbf M}]$ as well as the column matrix $[\bar{\mathbf u}_k]$, which are directly obtained from their respective counterparts without bar, and give the following linear eigenvalue problem
\begin{equation} 
	\Big[\big[ \, [\bar{\mathbf K}_d] + [\bar{\mathbf K}_s] \, \big] - \omega^2 [\bar{\mathbf M}] \Big]   [\bar{\mathbf u}_j] = [\mathbf 0] .
\end{equation}
The computed eigenfrequencies are the circular frequencies $\omega_m$ from which the frequencies $f_m$ are obtained by $\omega_m = 2 \pi f_m$ and the phase velocity by $c_{p,m} = \lambda f_m$.

This method was originally intended for periodic waveguides, but can also be used for homogeneous ones \cite{sorohan.2011,Gomez.2015}. It is effective and computationally efficient, especially for pre-stressed waveguides, since it can be employed directly with a general-purpose finite element code once the pre-stress state has been computed. In contrast to the semi-analytical finite element (SAFE) method \cite{Ahmad.2011,Galan.2002,Gao.2007,Treyssede.2010}, which assumes a displacement solution and requires the discretization of the cross-section of the waveguide,  the present method uses a geometric assumption of the waveguide, e.\,g. its periodicity, which is reflected by the boundary conditions \cite{Hakoda.2018}. 

\subsection{Numerical model and computational procedure}
\label{sec:NumMod}
To simulate the acoustoelastic effect in an isotropic waveguide, a 2D model is created representing the specimen in Figure~\ref{fig:Specimen}. The material properties of the numerical model are provided in Table~\ref{tab:MatProp}. The Young's modulus of the aluminum AlMg3 are experimentally obtained from the specimens used for the experimental investigation in Section~\nameref{sec:Experiments}. It is worth noting, that the Young's modulus depends slightly on the specimen's thickness. This will be considered when comparing the numerical and experimental results later on. The third-order elastic constants are taken from literature\cite{Stobbe.2005}. They are derived from acoustoelastic analyses based on the fatigue behavior of EN AW-7075. Even though this alloy differs slightly from the one used in the present study, this source is the best available in the literature for third-order elastic constants.

\begin{table}
\small\sf\centering
\caption{Material properties for the numerical simulation.}
\label{tab:MatProp}
\begin{tabular}{lcc}
\toprule
Parameter & Unit & Value\\
\midrule
E (2~mm) & GPa & 64.4 \\
E (0.5~mm) &  GPa & 68 \\
$\nu$   & - & 0.33 \\
$\rho$  & kg/m$^3$ & 2700\\
\midrule
$\ell$ & GPa & -255.2 \\
m & GPa & -325.0\\
n & GPa & -351.2\\
\bottomrule
\end{tabular}
\end{table}

\begin{figure}[hb]
  \centering
    \def\svgwidth{\columnwidth}
    \includegraphics{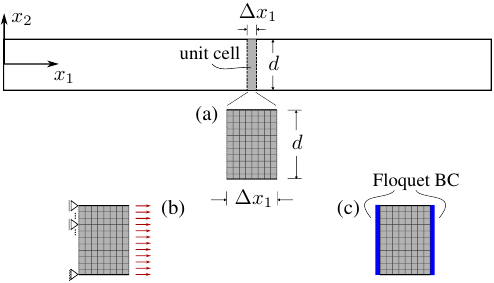}
    \caption{Scheme of specimen and disretized unit cell, not to scale (a), loading and boundary conditions for pre-stress analysis (b), Floquet boundary conditions (BC) for eigenvalue problem computation (c).}
  \label{fig:num_modell}
\end{figure}

The finite element model maps the $x_1$-$x_2$ symmetry plane. Therefore, the calculations are performed in the plane strain state. Also, the model includes the entire specimen thickness $d=$1.0~mm in $x_2$ direction. The periodic unit cell, which represents a small segment in $x_1$ direction in which the waves propagate, has the length $\Delta x_1 =\,$0.1~mm, see Figure \ref{fig:num_modell}~(a).

The discretization is carried out using finite elements with biquadratic shape functions and a length of 0.01~mm, so that a total of 1000 elements are used. In \cite{mace.2008,ichchou.2007}, it is recommended to use at least 6 elements per wavelength and 10 nodes per wavelength for high spatial resolution. This means that the finite element mesh is very fine in relation to the expected wavelength, even in the high-frequency range. However, the finite element mesh of the unit cell only has a small number of degrees of freedom, so that the computational time is short.

First, this finite element model is used to compute the pre-stress due to external loading in a standard nonlinear static analysis, see Figure \ref{fig:num_modell}~(b). Here, either the Neo-Hooke or Murnaghan material model is used. The boundary conditions correspond to those in a common tensile test. The results of this computation are the mechanical quantities of the intermediate configuration.

The subsequent computations simulate the wave propagation in the specimen and yield the mechanical quantities in the current configuration. As mentioned above, a linear eigenvalue problem is solved for this purpose. The left and right boundary conditions now reflect the Floquet periodicity and thus allow the calculation on a unit cell that repeats periodically in the $x_1$ direction. \cite{Hakoda.2018}, see Figure \ref{fig:num_modell}~(c). Dispersion curves at a given pre-stress are typical results of this analysis.

\subsection{Numerical results}

\begin{figure*}[thb]
    \centering
    \def\svgwidth{\columnwidth}
    \includegraphics{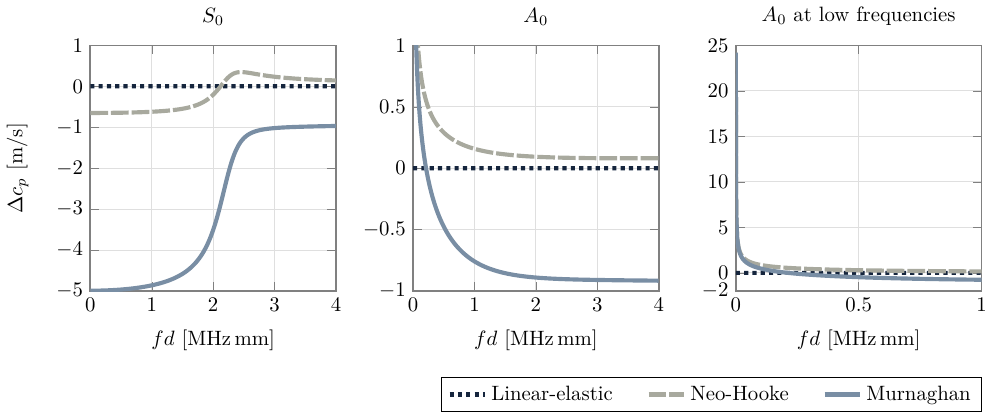}  
    \caption{Phase velocities differences $\Delta c_p$ of non-loaded and pre-stressed (100~MPa) specimens with different material models.}
    \label{fig:cp_diff_num}
\end{figure*}

Results of the numerical procedure for the characterization of the wave propagation in linear waveguides, especially the adapted eigenfrequency analysis to COMSOL Multiphysics,  can be found in previous work of the authors\cite{Mik.2022,Barth.GAMM.22.FML}. So, the focus of this subsection is on numerical investigations on pre-stressed structures.

Figure \ref{fig:cp_diff_num} shows a result of computations on the non-loaded and loaded strip-shaped specimen from Figure \ref{fig:num_modell}. The differences in the phase velocities when using a linear and both hyperelastic material models in the range of the frequency-thickness product from 0 to 4~MHzmm are shown. As expected, the linear material model makes no difference between non-loaded and loaded specimens. However, both hyperelastic material laws are able to represent phase velocities differences under load, which here is 100~MPa. It is worth noting that both the Murnaghan and Neo-Hooke curves have a zero crossing, albeit at different modes.

\begin{figure*}[htb]
  \centering
  \def\svgwidth{\columnwidth}
    \includegraphics{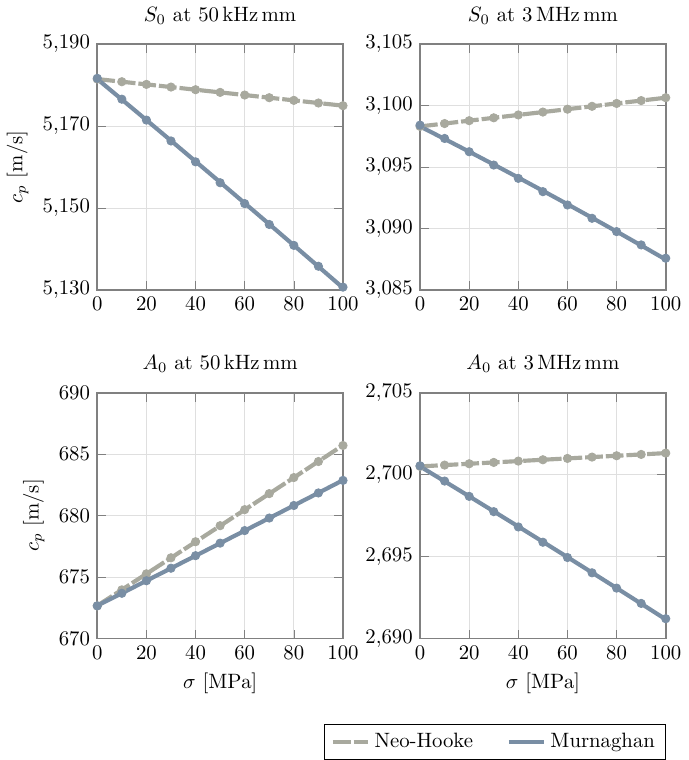} 
  \caption{Phase velocities with increasing pre-stress at 50~kHzmm (left) and 3~MHzmm (right) for the Neo-Hooke and the Murnaghan material model. Upper row: $S_0$ mode, lower row: $A_0$ mode.}
  \label{fig:lin_zus}
\end{figure*}

In addition, computations are carried out on strip-shaped specimens at increasing pre-stresses from 0 to 100~MPa. In Figure \ref{fig:lin_zus}, the results are presented for frequency-thickness products of 50~kHzmm and 3~MHzmm for both modes. It becomes obvious, that the results for the phase velocities at different pre-stress levels can be perfectly connected by a straight line for both material models and at both frequencies under consideration, so that a linear dependency can be concluded. Figure \ref{fig:lin_zus} also shows, that at the lower frequency, both material models have the same tendency, namely decreasing phase velocity with increasing pre-stress. However, at the higher frequency, the two material models show opposite behavior.

On the basis of these numerical results, it is not possible to assess which constitutive material model best reflects the physical phenomenon of acoustoelasticity. Therefore, a comparison with experimental results is made in the following.

\subsection{Comparison with experimental results}

\begin{figure*}[htb]
  \centering
  \def\svgwidth{\columnwidth}
    \includegraphics{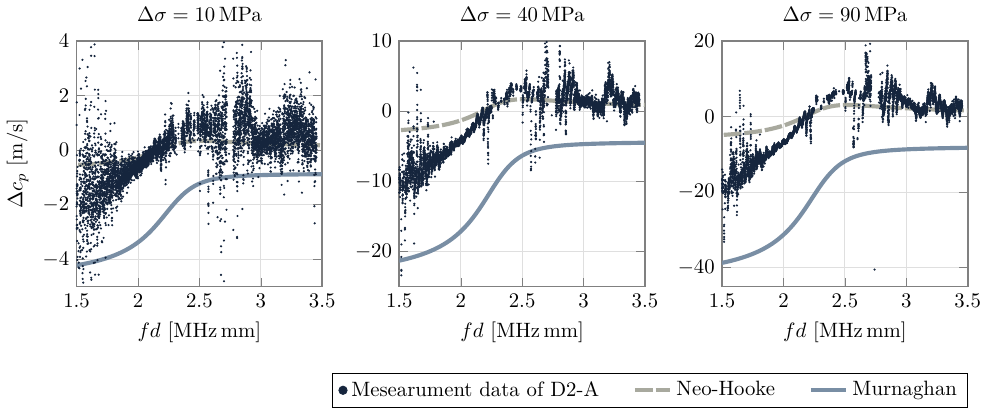} 
  \caption{Comparison of the phase velocity difference $\Delta c_p$ at different stress states based on the experimental and numerical data for the $S_0$ wave mode propagation in specimen D2-A.}
  \label{fig:cp_diff_exp_num_D2_S0}
\end{figure*}

In this section the numerically obtained results are finally compared with the experimental data presented above. The comparison follows the same scheme as done before for the numerical and experimental data, respectively. First, the phase velocity difference is observed for certain load levels over the whole frequency range covered during the experiments. This is followed by the analysis of the correlation between the pre-stress level and the phase velocity difference. 

\subsubsection{Phase velocity}

Figure~\ref{fig:cp_diff_exp_num_D2_S0} holds on the one hand the numerical results based on both the Neo-Hooke and Murnaghan material model as well as the experimental data for the $S_0$ wave mode based on the specimen D2-A. As discussed before in Section~\nameref{sec:NumMod} the Young's modulus for aluminum is set to 68~GPa, which corresponds to the values measured in preliminary tests for the specimens of thickness 2~mm.

Figure~\ref{fig:cp_diff_exp_num_D2_S0} reveals, that the numerical simulations based on the Neo-Hooke material model clearly reproduce the course of the measured data better than those based on the Murnaghan model. This applies here for all load steps considered. It is particularly noticeable in this comparison that both the measurement data and the Neo-Hooke data show a sign change in the phase velocity difference from negative to positive. This is not the case for the Murnaghan data. A more precise comparison shows that the Neo-Hooke data also shows a difference to the measured data. However, the comparison of the data curves is in a very good range.

\begin{figure*}[p]
  \centering
  \def\svgwidth{\columnwidth}
    \includegraphics{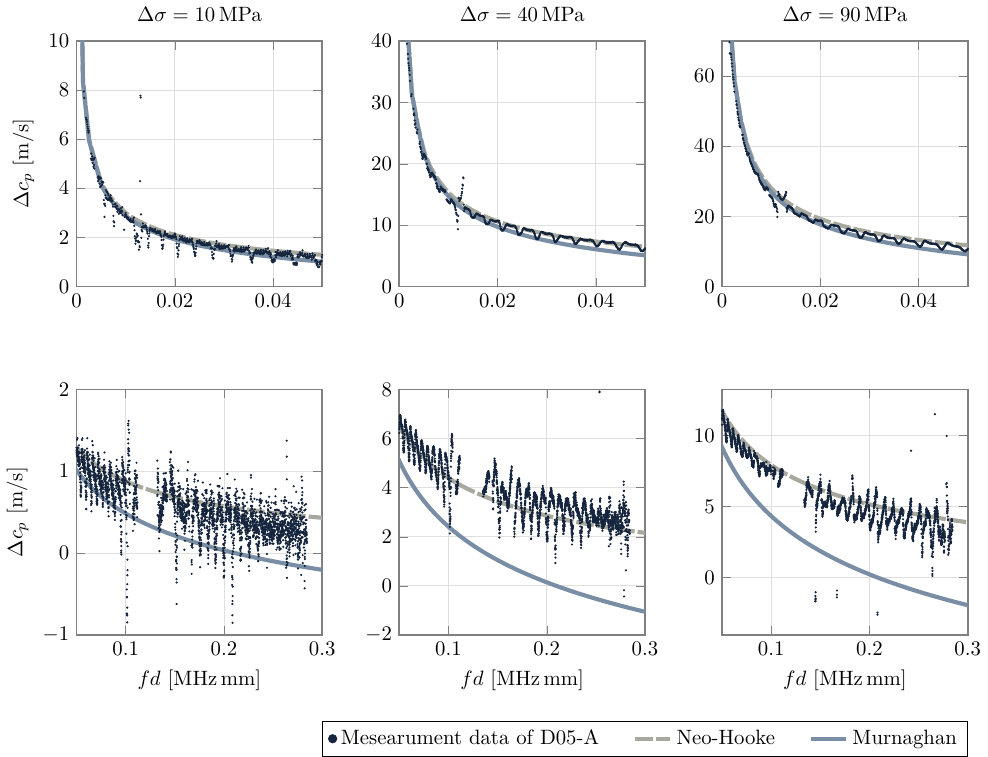} 
  \caption{Comparison of the phase velocity difference $\Delta c_p$ at different stress states based on the experimental and numerical data for the $A_0$ wave mode propagation in specimen D05-A.}
  \label{fig:cp_diff_exp_num_D05_A0_div}
\end{figure*}

Figure~\ref{fig:cp_diff_exp_num_D05_A0_div} shows the experimental data from the specimen D05-A and the numerical data for the $A_0$ mode in a frequency-thickness range of less than 280~kHzmm. In contrast to the comparison for the $S_0$ wave mode in this case a Young's modulus of 64.6~GPa is considered, see Section~\nameref{sec:NumMod}. For the sake of clarity, the data range is divided into two graphs. It is noticeable again, that the numerical data based on the Neo-Hooke material model shows a significantly better agreement with the measured data than those based on the material model according to Murnaghan. In contrast to the comparison for the $S_0$ wave mode, the Murnaghan model predicts a sign change for the phase velocity difference in this case. However, this prediction is not supported by both the measurement data and the numerical simulation utilizing the Neo-Hooke material model. Compared to the observation of the $S_0$ wave mode in Figure~\ref{fig:cp_diff_exp_num_D2_S0}, the data according to Neo-Hooke material model once again better represent the course of the measurement data. 

\begin{figure*}[p]
  \centering
  \def\svgwidth{\columnwidth}
    \includegraphics{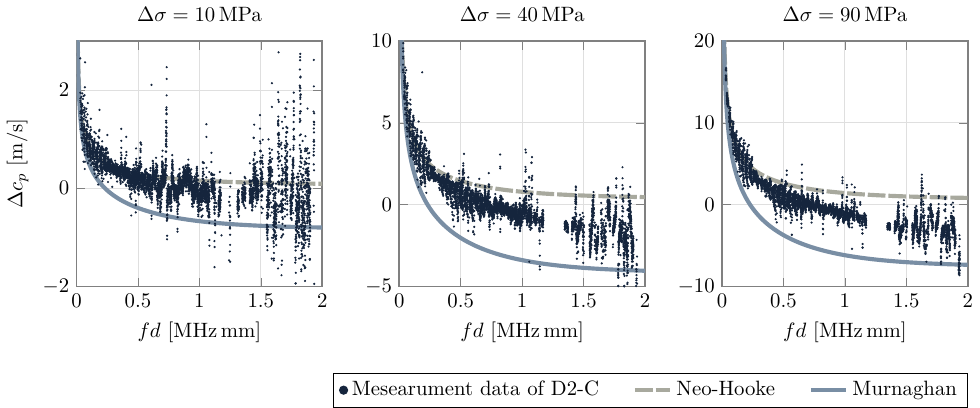} 
  \caption{Comparison of the phase velocity difference $\Delta c_p$ at different stress states based on the experimental and numerical data for the $A_0$ wave mode propagation in specimen D2-C.}
  \label{fig:cp_diff_exp_num_D2_A0_div}
\end{figure*}

In Figure~\ref{fig:cp_diff_exp_num_D2_A0_div}, the comparison of the $A_0$ wave mode is extended to a frequency-thickness range of up to 2~MHzmm. In contrast to the previous consideration, the numerical data are computed with a Young's modulus of 68~GPa. The diagrams show that in the low frequency-thickness product range below approximately 0.4~MHzmm, the Neo-Hooke material model approximates the measurement data much better than the Murnaghan model. However, above this frequency-thickness product the measured data starts to deviate from the numerical results. Furthermore, Figure~\ref{fig:cp_diff_exp_num_D2_A0_div} reveals again the contradiction between the Murnaghan material model and the Neo-Hooke material model regarding a sign change of the phase velocity difference. Based on the Neo-Hooke material model there is no sign change, whereas the Murnaghan material model still predicts it at approximately 0.2~MHzmm. A comparison with the experimentally obtained data appears to be inconclusive at this stage, since the data shows a sign change at approximately 0.6~MHzmm. 

Following this it is concluded, that for the $A_0$ wave mode the low frequency-thickness range below approximately 0.4~MHzmm is well represented by a numerical simulation based on the Neo-Hooke material model. For higher frequency-thickness products, however, the data shows significant deviations including a missing sign change of the phase velocity difference. The Murnaghan model, on the other hand, provides a greater deviation from the experimental data over the entire considered frequency-thickness range. In contrast to the Neo-Hooke material model a sign change is predicted, which meets the experimental data. Although the position of this sign change differs significantly.

\subsubsection{Stress}

\begin{figure*}[htb]
  \centering
  \def\svgwidth{\columnwidth}
    \includegraphics{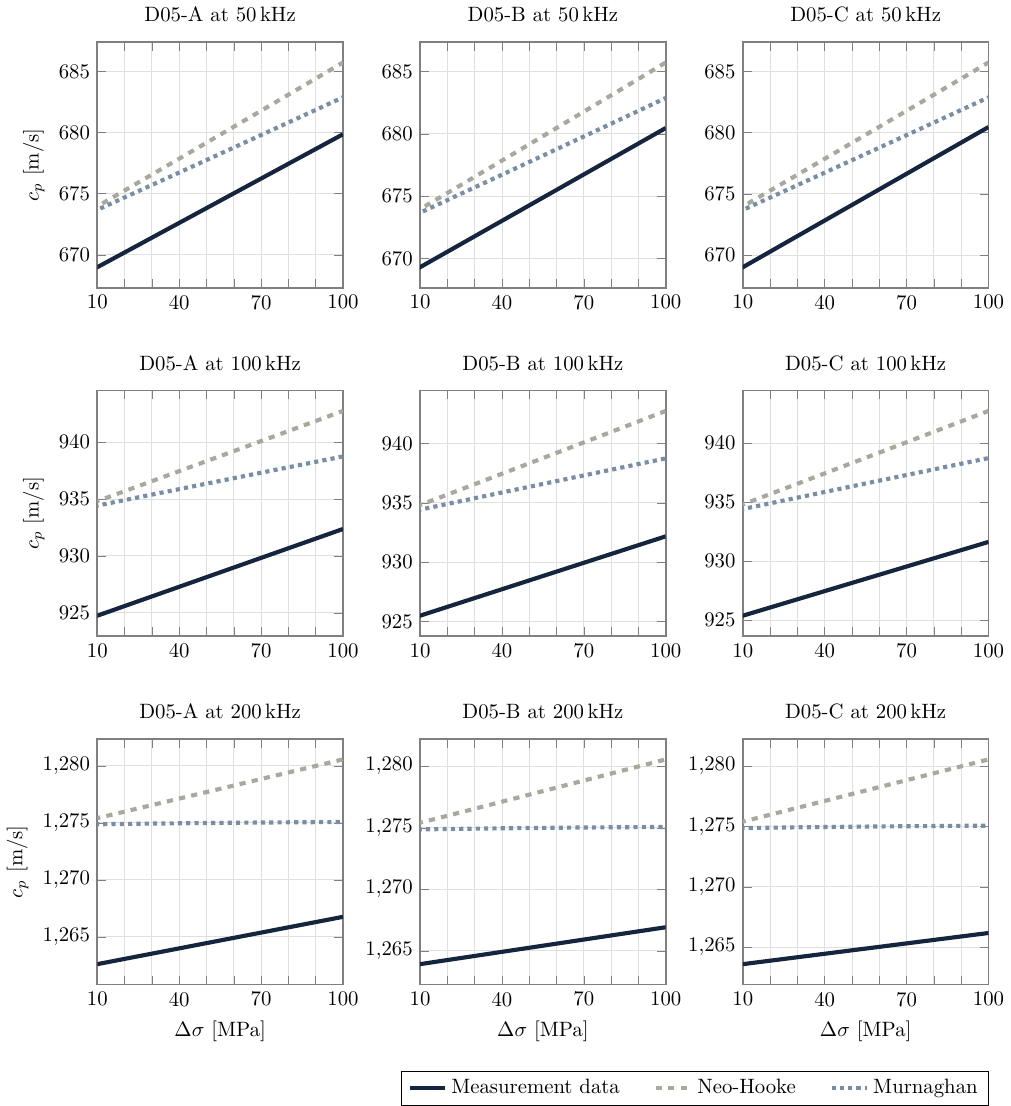} 
  \caption{Comparison of the linear regression based on the measurement data for the three specimens D05-A, D05-B, and D05-C with results derived from numerical simulation taking into account the Neo-Hooke and Murnaghan material model.}
  \label{fig:Linear_Verlauf_vergleich_exp_num}
\end{figure*}

As already discussed in the context of the experimental data, the comparison is extended to the linear regression of the phase velocity as a function of the pre-stress state difference. Due to the quality of the measurement data and the derived phase velocity differences, this comparison is limited to the $A_0$ wave mode. Figure~\ref{fig:Linear_Verlauf_vergleich_exp_num} shows the regressions based on the experimental and numerical data for frequencies of 50~kHz, 100~kHz, and 200~kHz, which are identical to the  frequencies depicted in Figure~\ref{fig:Linear_Verlauf_exp}. 

The comparison meets the expectations. The slopes of the numerical data based on the Neo-Hooke material model fit very well to those from the experimental data. For the Murnaghan material model, on the other hand, the slope differs significantly. However, due to a variation of the material properties between the numerical simulation and the specimens the initial phase velocity are not the same. 

\begin{figure*}[htb]
  \centering
  \def\svgwidth{\columnwidth}
    \includegraphics{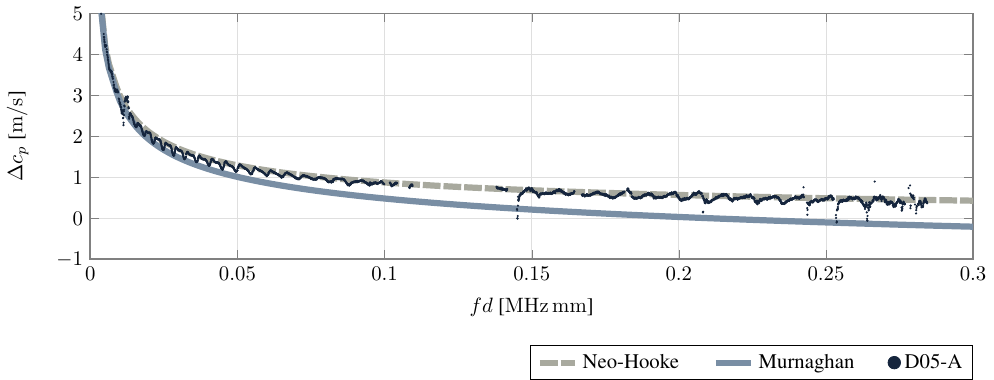} 
  \caption{Comparison of the computed unity uniaxial load step of 10~MPa for both the Neo-Hooke and Murnaghan material model with the experimental data of the sprecimen D05-A.}
  \label{fig:Einheitsschritt}
\end{figure*}

To depict the conclusion of this work in a more comprehensive way, the mean phase velocity difference for a load step of 10~MPa is computed for each frequency within the range of up to 0.3~MHzmm from the linear regression. Since the obtained value comprises the information of all analyzed load steps it can be interpreted as the phase velocity difference due to a unity uniaxial load step of 10~MPa. The results are summarized in Figure~\ref{fig:Einheitsschritt} and reveal that the Neo-Hooke material model is very suitable for the representation of the acoustoelastic effect in a frequency range of up to 0.3~MHzmm.

\section{Summary and conclusion}
This work deals with both numerical and experimental methods for investigations of acoustoelastic problems. The methods are exemplarily applied to GUWs in pre-stressed aluminum waveguides. The results are presented and discussed and serve to verify the numerical model.

First, the continuum mechanical fundamentals of wave propagation in pre-stressed structures are discussed. The material behavior is assumed to be hyperelastic and the Murnaghan material law, which is often applied to acoustoelastic problems, and the Neo-Hooke material model, which is broadly used and less complicated to handle, are proposed for the constitutive description.

The aim of the experimental investigations is to determine the phase velocities of the fundamental wave modes $S_0$ and $A_0$. For this purpose, non-loaded specimens and those loaded in a test machine with stepwise increasing pre-load are examined. The wave propagation is captured with a scanning laser vibrometer. A special multi-frequency excitation signal enables the analysis of the phase velocities of propagating waves in a wide frequency range. This considerably expands the data available in the literature.

The numerical analysis is carried out using the finite element method. First, the pre-stress in the specimen is calculated in a non-linear computation. The associated phase velocities of propagating waves are determined as a result of an eigenvalue analysis on a small unit cell of the specimen. The resulting dispersion diagram is valid for the previously determined pre-stress. This procedure is computationally very effective and has the advantage that it can be carried out with standard finite element programs. Therefore, it is an appropriate and useful alternative to the SAFE method, which is frequently used for this purpose.

A number of results first show that the experimental and numerical methods proposed in this work provide plausible results. It is also demonstrated that the numerical and experimental results are in very good agreement with the experimental data. Even though individual experimental observations are not exactly reproduced, the comparison shows on the whole that the Neo-Hooke material model approximates the experimental results better than the Murnaghan model. This result is significant because the Neo-Hooke model requires fewer material parameters than the Murnaghan model, is therefore easier to handle and is available in most finite element program systems.

Future work will deal with layered and complex material systems, e.\,g. fiber metal laminates, more complex pre-stressing states as well as residual stresses.

\begin{funding}
The work was funded within the Research Unit 3022 ‘Ultrasonic Monitoring of Fibre Metal Laminates Using Integrated Sensors’ by the German Research Foundation (Deutsche Forschungsgemeinschaft (DFG)).
\end{funding}

\begin{acks}
The authors expressly acknowledge the financial support for the research work on this article within the Research Unit 3022 ‘Ultrasonic Monitoring of Fibre Metal Laminates Using Integrated Sensors’ by the German Research Foundation (Deutsche Forschungsgemeinschaft (DFG)).
\end{acks}

\begin{dci}
The authors declared no potential conflicts of interest with respect to the research, authorship, and/or publication of this article.
\end{dci}

\bibliographystyle{SageV}

\end{document}